Comments on Letter (Phys. Rev. L, Vol. 89, No. 10, 2002) by D. Shapira & M. Saltmarsh

R.P. Taleyarkhan[1], C.D. West[2], J. Cho[3], R.T. Lahey Jr.[4], R.I. Nigmatulin[5], R.C. Block[6]*
(1) – Purdue University, W. Lafayette, IN 47905, USA; Email: rusi@purdue.edu
(2) – Univ. Tennessee, Knoxville, TN, USA; Email: herderwest@comcast.net
(3) – FNC Technology Co., Soeul, S. Korea 151-742;Email: chojs@fnctech.com
(4) Rensselaer Polytechnic Institute, Troy, NY, 12180, USA; Email: laheyr@gmail.com
(5) –Russian Academy of Sciences, Inst. Oceanology, Moscow, Russia; Email: nigmar@ocean.ru
(*) – corresponding author: blockr@rpi.edu

This article focuses on correcting several factual errors and critiques in the previously published Letter [1] by Shapira and Saltmarsh. In their Letter [1], the authors claim, "We have repeated the experiment of Taleyarkhan et al. [2]" This is incorrect. The published data provided by the authors of the Letter [1] did not come from their own independently performed experiment; these data came from an experiment that we performed, on our experimental apparatus, in our laboratory. The authors of the Letter [1] were observers who brought in their own neutron detector and obtained data on a single day, for two one-hour periods (the first with cavitation on, immediately followed by another with cavitation off).

The authors of the Letter [1] failed to make a proper comparison between neutron emission and tritium production. The correct protocol, as shown in our claims [2], is to perform multiple separate experiments (for cavitation on followed by cavitation off) in which assessments of neutron emission and tritium production are conducted concurrently. The protocol includes not only performing experiments using deuterated acetone ($C_3D_6O$) but also performing similar control experiments using normal acetone ($C_3H_6O$). The authors of the Letter [1] did not perform control experiments with $C_3H_6O$ [3], and, more important, they did not monitor for tritium production.

The claim by the authors of the Letter [1] of a mismatch of neutron-to-tritium production is invalid because data must be taken from experiments in which neutron emission and tritium emission measurements are made concurrently from a single experiment. The authors of the Letter [1] used measured neutron data from some other experiment, on some other day, and compared that with our tritium data taken from a different set of experiments on other days. Using these data, they depicted a ratio of the neutron emission rate to the tritium emission rate for the experiment they observed. This cannot be done.

In their Letter [1], the authors failed to disclose that the data which they collected from our experiment confirm our claims. We have communicated [4] to the authors of the Letter [1] that the raw data that they collected while observing our experiment reveal statistically significant (over 9 SD) neutron emissions during the first period (30-150 μs), which is time-correlated with the first set of bubble implosions and sonoluminescence (SL) emission. After pulsed neutron generator (PNG) firing, a statistically significant (~ 7 SD) excess of nuclear emissions also appears over background, which is time-correlated with the second span of SL emissions during the second period (700-to-1700 μs), after PNG neutron burst emission. We present the results of our assessments in Figs. 1 and 2 of the attached supplement [3] and elsewhere [2,5].

The authors of the Letter [1] speculate that the statistically significant neutron emissions are caused by room-return neutrons. This explanation fails. The net effect of room-return neutrons from cavitating vs. non-cavitating experiments using $C_3D_6O$ should be null because the experimental configurations and apparatus remained the same between their two one-hour data acquisition runs. The first run was performed with cavitation on, the next one with cavitation off. The data acquired by the authors of the Letter [1] display a distinct time structure in the microsecond frame during each 5ms sweep (see Figs. 1 and 2 of Ref. 3 to this Comment).

In summary, we have shown that the measurements by the authors of the Letter [1] did, indeed, reveal a statistically significant (over 7 to 10 SD) emission of nuclear (neutron-gamma) emissions when the system, consisting of chilled deuterated acetone, was placed in the mode of cavitation. Furthermore, we have shown that the conjecture in the Letter [1] of room-return background is incorrect. We also critique the Letter [1] by pointing out that its authors failed to perform concurrent tritium production measurements. Therefore, the authors' claim of a mismatch between neutron and tritium emission rates is without basis.

We also caution in relation to the fact that monitoring for possible tritium generation during bubble fusion experiments must be conducted without significant delay in time (i.e., within hours of completion); this is an important consideration in order to avoid loss/egress of tritium as has been ascertained via specifically scoped experiments[6] using deliberately injected tritium in quantities similar to those we have reported from chilled deuterated acetone (but instead of generation, the tritium was obtained from a standard supply); it was revealed that a lapse of close to a day or more resulted in sufficient loss of tritium and liquid mass from the scintillation vials such that initially positive indications for the presence of tritium, then, after a time span of over 20h resulted in null results. For our bubble fusion experiments[2] we conducted tritium monitoring within 2-3 hours of completion of experimentation; as was also the case for successful bubble fusion experiments reported by others.[7,8] On the other hand, the results of scoping experiments of Tsoukalas et al.[9] reported null results. From personal communications with individuals involved with that effort[10] it was revealed that while initial monitoring (within hours of completion) did indeed reveal statistically significant tritium production, the published results reported in the archival literature were obtained from their later measurement-conducted after several weeks to months of completion of their experiments; however, this was not clarified for this aspect.

**Acknowledgments:**
The authors are grateful to Shapira and Saltmarsh for their cooperation and to Shapira for his kind transmittal of their raw data, for enabling our independent analyses.

**References:**
1. D. Shapira and M. Saltmarsh, *Phys. Rev. L,* Vol.89, No.10,104302-1, (2002).

# Ref. 3 - Supplement to Article:

In this supplement we present results of our assessments of the data obtained by Shapira and Saltmarsh in 2001 and also, to point to critical differences between our reported evidence for nuclear emissions during acoustic cavitation [2] and that reported by Shapira and Saltmarsh in their Letter [1].  We summarize this information related to both neutron and tritium emissions we have reported upon in our 2002 *Science* article.

**Our Assessments of Raw Data Obtained by Shapira and Saltmarsh [3]**

 We have independently assessed the raw data kindly supplied [3] by the authors of the Letter [1] and find that in addition to statistically significant (over 9 SD) neutron emissions during the first period (30-150 µs) time correlated with first set of bubble implosions and sonoluminescence (SL) emission, after PNG firing, there appears a statistically-significant (~ 7 SD) excess of nuclear emissions over background time correlated with the second span of SL emissions between the time period 700-to-1700 µs after PNG neutron burst emission.  The results of our assessments are presented in Figs. 1 and 2.

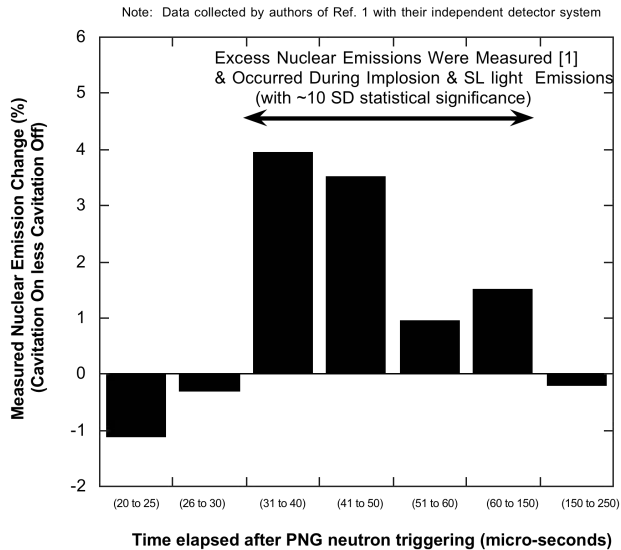

Figure 1.  Measured excess nuclear emissions with cavitation on vs off in deuterated acetone for region of time corresponding to first collapse period of PNG neutron nucleated bubbles – from raw data transmitted [4] by authors of Letter [1].

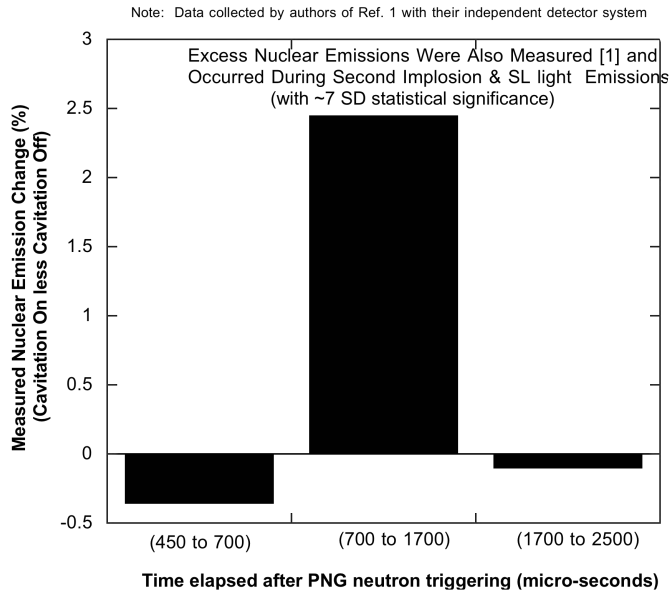

Figure 2. Measured excess nuclear emissions with cavitation on vs off in deuterated acetone for region of time corresponding to second subsequent collapses of PNG neutron nucleated bubbles) – Data taken with PD detection system by authors of Letter [1]

**Comments on Our Neutron Emission Related Work for *Science* [2]**

Colin D West (CDW) had retained his notes of 12/15/2001 related to neutron emission related experiments conducted in our laboratory at Oak Ridge National Laboratory (ORNL). These data were obtained with an Elscint™ liquid scintillation (LS) detector borrowed from Rensselaer Polytechnic Institute (RPI) and set up/calibrated at ORNL for bubble fusion data acquisition by co-author (Robert Block). Table 1 summarizes these data from handwritten worksheet.

| Table 1. Summary of data on neutron counting rates[a] found on a CDW handwritten worksheet dated 12-15-01 | | | |
|---|---|---|---|
| # | Counting Period (s) | Number of counts measured[b] During Cavitation | Without Cavitation |
| 1 | 100 | 15,115 | 14,566 |
| 2 | 100 | n.a[c] | 14,251 |
| 3 | 300 | 47,220 | 45,488 |
| 4 | 300 | 47,226 | 44,916 |
| 5 | n.a[d] | n.a[d] | n.a[d] |
| 6 | 300 | 51,004 | 49,616 |

    (a) Neutrons detected, by the Escint™ with pulse-shape discrimination (PSD) system installed by R.C.Block, during PNG operation
    (b) Counts in channels at or below the 2.45 MeV proton recoil edge (PRE)
    (c) This entry was blank on CDWs 12-15-01 worksheet
    (d) CDWs worksheet notes that, when copying down the information, sheet 5 was was not readily noted for documentation on that occasion.

The procedure for conducting tests were such that the run with cavitation on for a given period of time was done first and then followed with a run with cavitation off for the same period with experimentation staff and other parameters remaining the same. Per recollection of protocols followed during 2001, the gap in between the cavitation on and cavitation off runs would nominally be less than a minute.

Using the data from Table 1, one sees a total of 160,565 counts in four separate runs over a combined 1,000 seconds of cavitation. Using the Poisson standard deviation (S.D.) only, that corresponds to 160.57 +/- 0.40 counts per second (cps). Likewise, the five separate runs with no cavitation total 168, 837 counts in 1,100 seconds, or 153.49 +/- 0.37 cps.

<u>The difference (increase) during cavitation is 7.08 +/- 0.55 cps, which is more than 12 S.Ds above zero.</u>

Table 2, below, shows the number of counts; the Poisson S.D. of that number; and the corresponding counting rate for each of the results listed in Table 1. We note that in Table 2, the differences among successive "during" cavitation runs are much larger than their Poisson S.Ds, and likewise for the "without" cavitation runs.

| Table 2. Aggregating the Elscint/PSD Neutron Counts Below the 2.45 MeV PRE | | | | | | | |
|---|---|---|---|---|---|---|---|
| # | Counting time (s) | During Cavitation | | | Without Cavitation | | |
| | | Counts | S.D. | CPS | Counts | S.D. | CPS |
| 1 | 100 | 15,115 | 123 | 151.2 +/-1.2 | 14,566 | 121 | 145.7 +/-1.2 |
| 2 | 100 | n.a. | | | 14,251 | 119 | 142.5 +/-1.2 |
| 3 | 300 | 47,220 | 217 | 157.4 +/-0.7 | 45,488 | 213 | 151.6 +/-0.7 |
| 4 | 300 | 47,226 | 217 | 157.4 +/-0.7 | 44,916 | 212 | 149.7 +/-0.7 |
| 6 | 300 | 51,004 | 225 | 170.0 +/-0.8 | 49,616 | 223 | 165.4 +/-0.7 |

We believe that the implication is that there are real variations in the counting rates on a time scale that is longer than the duration of the "with" and "without" cavitation experiment pairs, i.e., longer than ~600 seconds. Thus, both numbers of each pair could come from similar, normally distributed populations but conditions may have changed before the next pair of measurements were taken.

Possible causes for such variations include changes in the liquid temperature (which would, for example, change the cavitation threshold and hence the volume of liquid in which bubbles can be nucleated and also the fraction of the acoustic period during which the negative pressure is greater than the cavitation threshold). Also, according to our calculations the D-D fusion reaction rate is extremely sensitive to the liquid temperature. Alternately, the high gain electronic components may give rise to varying efficiencies for detection over several hours of operation, due possibly to thermal issues, etc. Or perhaps the tuning of the piezoelectric drive frequency to the chamber resonant frequency, which depends on the temperature and the exact liquid level, was not always exactly the same. Additional statistical assessments were also made by us using methods from the excellent book by J. Toppings "Errors of Observation and Their Treatment," 4[th] Ed., Chapman and Hall, 1972. This provided an estimate indicating that systematic changes larger than Poisson S.D.s are present and that the weighted mean of observations of increases in neutron count rate during cavitation (~6.01 +/-0.7 cps), is more than eight standard errors above zero.

At this point, it is worth noting again that Shapira and Saltmarsh had not, in fact, ".. repeated the experiment of Taleyarkhan et al.." They did not report any short counting runs but just run with cavitation on and one measurement with cavitation off (each of about an hour's duration) which would have masked the differences in count rates that can be clearly seen in Table 1 results from multiple pairs of short tests made in quick succession.

This means that Shapira and Saltmarsh did not "repeat" a single one of our experiments. They measured the neutron count changes in a different experiment, of their own devising, that did not avoid systematic long-term changes in the performances of the acoustic chamber. Shapira and Saltmarsh did indeed find, despite having conducted a different experiment of their devising, 7 to 10 S.D. increase in nuclear emissions, which are time correlated with the region of time corresponding with bubble implosion and sonoluminescence (SL) light production. Shapira and Saltmarsh did not make any tritium measurements – they just made neutron yield measurements – and thus, they cannot compare with our tritium measurements.

**Comments on Our Tritium Emission Related Work for *Science* [2]**

We have already reported our tritium emission data graphically [2]. An assessment of these data result in the conclusion that on the aggregate, excess tritium emissions with cavitation in deuterated acetone amounts to 7 S.Ds in statistical standard errors above zeroes for the 5 observations listed therein. Once again, we note that Shapira and Saltmarsh did not perform tritium emission experimentation. Therefore, we do not have a basis for comparing our tritium emission data with the neutron emission data procured by Shapira and Saltmarsh

**Comparing Measured Neutron and Tritium Emission Rates**

It is often taken for granted that the neutron and tritium emission rates from D-D thermonuclear fusion should be equal. However, this assumption is predicated on the plasma state being at a certain temperature above which the neutron branch is preferred, and below that a preference is towards more tritium vs neutron emission. This is noted from Fig. 3 (Curtiss, 1959). As we note, from the low end energy of 20 keV to the upper energy levels of ~300 keV for the bombarding deuterons, the neutron-to-tritium branching ratio can vary from 0.9 to over 1.15 (i.e., a spread of roughly 25%].

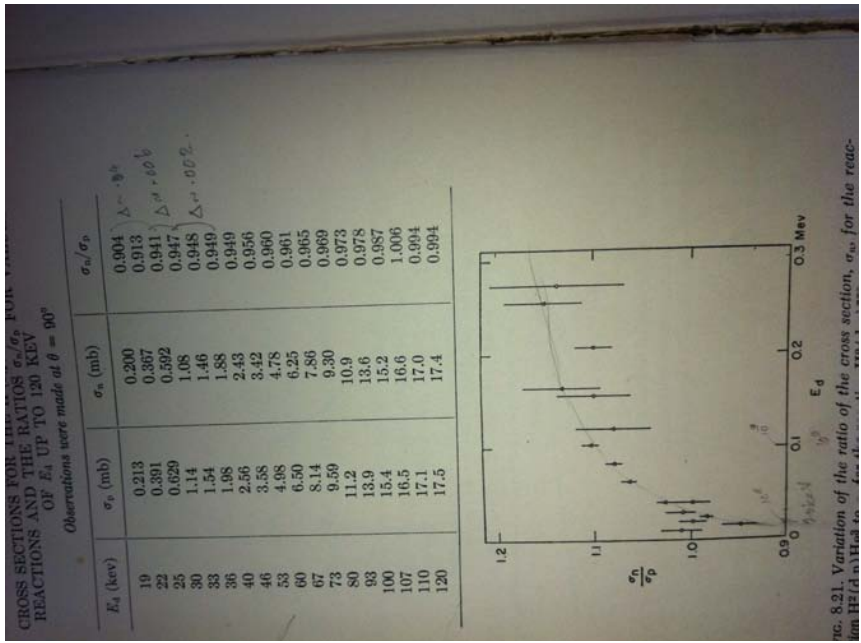

Figure 3. Variation of the ratio of the cross-section for the reaction $^2$H(d,n)$^3$He to the cross-section for the reaction $^2$H(d,p)$^3$H with the energy of the bombarding deuteron (page 118 from Curtiss, 1959).

Other uncertainties must also be taken into account for bubble fusion experimentation in which neutron data were obtained over short time periods of less than an hour versus tritium data, being required (due to instrument limitations) to be acquired over 7 to 12 hours duration. The Beckman™ LS6500 spectrometer was utilized for tritium monitoring for which we estimated that the counting efficiency was about 70%.

Other sources of uncertainty that we must keep in mind are provided below along with their respective impacts on estimation of either neutron or tritium emission rates:

- <u>The distribution of excess tritium produced during acoustic cavitation.</u> The 1 mL samples of deuterated acetone were taken from the top regions where most of the cavitation bubbles were nucleated. By assuming that the measured tritium amount would be representative for the entire 330cc volume of the chamber provides an upper bound. We made calculations for 50% and 100% of the chamber volume. Values of triton (T) emission rate for each of the 5 runs with chilled cavitated deuterated acetone are presented in Table 3. As noted therein, our estimated tritium production rate on average ranged from ~5x10$^5$ T/s to ~7.9x10$^5$ T/s. These estimates would double if we assume that the entire 330 cc vs half of the volume was involved.

- <u>Calibration uncertainties of the Elscint™ detector.</u> For estimating the intrinsic efficiency of downscattered 2.45 MeV neutron counting we had used a 1 Curie Pu-Be isotope source (originally certified in the 1960s timeframe to provide a neutron production intensity of about $2 \times 10^6$ n/s). However, we must note that a Pu-Be source provides neutrons of energies from 0.1 MeV through 11 MeV. Assuming an efficiency for the neutron range of 0.8 MeV (the lower level setting of our detector) through only 2.45 MeV to be the same as for neutrons of much higher efficiencies implies an overestimate (we estimate at least ~50%) for the efficiency. During transit of 2.45 MeV neutrons from within the chamber through acetone and glass before reaching the detector also means losses (we estimate ~50%) of neutrons outside of the 0.8MeV to 2.45MeV window. Finally, one also must account for a little known fact that a Pu-Be neutron source (with $^{239}$Pu as the main ~5.1MeV alpha emitting isotope) also may contain varying levels of $^{241}$Pu during the production stage. $^{241}$Pu is a beta emitter with a 14.5y half-life, which then decays to $^{241}$Am (with ~458y half-life); $^{241}$Am is a 5.5 MeV alpha emitter. Since a Pu-Be source produces neutrons from alpha particle interactions with Be, the neutron output over time actually increases on the net when the source is used over several decades. Knolls (2000) provides an estimate of a ~2% initial increase per year if the Pu-Be source contains ~0.7% of $^{241}$Pu. Of-course, the rate of increase can not keep remaining the same because the $^{241}$Pu begins to deplete itself simultaneously; an equilibrium of about 35% to 40% increase in neutron intensity is reached in about 20-30 years from start (if one assumed 0.7% as the initial $^{241}$Pu baseline) must be accounted for since the Pu-Be sources used in our experimentation during 2001 were originally produced in the 1960s. For this reason, we deem another factor of ~2 to be reasonable to use (the actual $^{241}$Pu content may have been greater or smaller than 1.5% but we do not have certification details or knowledge of the precise feedstock and nuclear reactor where the Pu was derived from; also, a typical nuclear power reactor at the end of cycle produces Pu isotopes within which the $^{241}$Pu fraction could be as high as 15%; hence considerable uncertainty can develop over several decades). Nevertheless, one must accept the fact that neutron measurements can be tricky and will encompass a level of uncertainty. Using only a raw count rate with PSD in place and using the Pu-Be source, we had estimated (including the solid angle effect) a counting efficiency ranging from ~ 1 to $2 \times 10^{-4}$. If we take into account the above-mentioned additional uncertainties this value of efficiency for counting drops to ~$1.2 \times 10^{-5}$. The measured neutron emission rates (based on experiments we conducted) then range from ~$4 \times 10^5$ n/s to ~$6.4 \times 10^5$ n/s at the upper end.

| Table 3. Tritium production rate estimation from data of Ref. 2. | | |
|---|---|---|
| Time duration for cavitation (hours) | Measured excess tritium count rate from 1 mL of sample (cpm) | Estimated "averaged" rate of tritium production over time of cavitation (T/s) |
| 7 | 5.75 | $5.05 \times 10^5$ |
| 7 | 6 | $5.26 \times 10^5$ |
| 7 | 7 | $6.14 \times 10^5$ |
| 12 | 13 | $6.65 \times 10^5$ |
| 12 | 15.5 | $7.93 \times 10^5$ |

Notes:
- Tritium measurement efficiency = 70%
- Volume for excess tritium bearing deuterated acetone in calculations = 166 mL
- Tritium production rate is assumed to be constant over duration of cavitation.
- Assumed half-life for tritium decay = 12.3 years

In a companion theory-based paper we have published in *Physics of Fluids* journal (Nigmatulin et al., 2005) the modeling framework when applied to our experiments reported in Science [2] offered an estimate of neutron production intensity of ~$5 \times 10^5$ n/s.

Table 4 summarizes the best estimates of tritium and neutron production during our experiments. As noted therein, the measured neutron and tritium estimates are close to each other and both are commensurate and in harmony with theory-based expectations.

| Table 4. Comparison of Estimates for Neutron and Tritium Emission Rates Between Measurements and Theoretical Predictions | | |
|---|---|---|
| Excess Neutron Emission Rates (n/s) - Measured[a] | Excess Tritium Emission Rates (T/s) – Measured[b] | Predicted[c] Neutron or Tritium Rates (n/s;T/s) |
| $4 \times 10^5$ to $6.4 \times 10^5$ | $5 \times 10^5$ to $7.9 \times 10^5$ | ~$5 \times 10^5$ |

a – Assuming $^{241}$Pu at 1.5% concentration; detection efficiency ~$1.25 \times 10^{-5}$
b – Assumed measured tritium amount represents production in 50% of volume and 70% LS6500 spectrometer efficiency.
c – Nigmatulin et al. (2005)

**References Cited from Comment**

1. D. Shapira and M. Saltmarsh, *Phys. Rev. L,* Vol.89, No.10,104302-1, (2002).
2. R. P. Taleyarkhan, C. D. West, J.S.Cho, R. T. Lahey,Jr., R. I. Nigmatulin, and R. C. Block, *Science* 295, 1968 (2002)-Supplement/Ref.32.
3. Personal communication to R. P. Taleyarkhan from D. Shapira (2001).

**Additional References Cited in Supplement:**

- Curtiss, L. F., "Introduction to Neutron Physics," D. Van Nostrand Company, Inc., Princeton, New Jersey, U.S.A., 1959.
- Knolls, G. F., "Radiation Detection and Measurements," 3rd Ed., John Wiley and Sons, Inc., 2000.
- Nigmatulin, R. I., I. Akhatov, A. Topolnikov, R. Bolotnova, N. Vakhitova, R. T. Lahey, Jr., and R. P. Taleyarkhan, "Theory of Supercompression of Vapor Bubbles and Nanoscale Thermonuclear Fusion," *Physics of Fluids* 17, 107106, American Institute of Physics, 2005.
- Toppings, J., "Errors of Observation and Their Treatment," 4th Ed., Chapman and Hall, 1972.